\documentclass[aps,showpacs]{revtex4}

\usepackage{epsfig}

\begin{document}

\title{Characterization of fragment emission in $^{20}$Ne (7 - 10 MeV/nucleon) + $^{12}$C reactions}

\author {Aparajita Dey$^{1}$, C.~Bhattacharya$^{1}$, S.~Bhattacharya$^{1}$, S.~Kundu$^{1}$, K.~Banerjee$^{1}$, \\ S.~Mukhopadhyay$^{1}$,
D.~Gupta$^{1}$, T.~Bhattacharjee$^{1}$, S.~R.~Banerjee$^{1}$, S.~Bhattacharyya$^{1}$, \\ T.~K.~Rana$^{1}$, S.~K.~Basu$^{1}$, R.~Saha$^{1}$, K.~Krishan$^{1}$\footnote {Present address : 306, VIP Enclave, VIP Road, Kolkata - 700 059, India}, A.~Mukherjee$^{1}$\footnote {Present address : Saha Institute of
Nuclear Physics, 1/AF Bidhan Nagar, Kolkata - 700 064, India.}, D.~Bandopadhyay$^{1}$\footnote {Present address : TRIUMF,
4004 Westbrook Mall, Vancouver, Canada V6T2A3.}, C.~Beck$^{2}$}

\affiliation{$^{1}$ Variable Energy Cyclotron Centre, Sector - 1, Block - AF,  Bidhan Nagar, Kolkata - 700 064, India. \\
$^{2}$ Institut Pluridisciplinaire Hubert Curien, UMR7500, CNRS-IN2P3 et Universite Louis Pasteur, \\
23, Rue du Loess, B.P. 28, F-67037, Strasbourg Cedex 2, France.}

\begin{abstract}
The inclusive energy distributions of the complex fragments (3
$\leq$ Z $\leq$ 7) emitted from the bombardment of $^{12}$C by
$^{20}$Ne beams with incident energies between 145 and 200 MeV have
been measured in the angular range 10$^{o} \leq \theta_{lab} \leq$
50$^{o}$. Damped fragment yields in all the cases have been found to
have the characteristic of emission from fully energy equilibrated
composites. The binary fragment yields are compared with the
standard statistical model predictions. Whereas Li and Be fragments yields are in
agreement with statistical-model calculations, enhanced yields of 
entrance channel fragments (5 $\leq$ Z $\leq$ 7) indicate the survival of
orbiting-like process in $^{20}$Ne + $^{12}$C system at these
energies.
\end{abstract}

\pacs{25.70.Jj, 24.60.Dr, 25.70.Lm}

\maketitle

\section{Introduction}
Several experiments have been done in recent years to understand the
mechanism of complex fragment emission in low-energy (E$_{lab}$
$\lesssim$ 10 MeV/nucleon) light heavy-ion (A$_{projectile}$ +
A$_{target}$ $\lesssim$ 60) reactions
\cite{Sanders99,carlin,pada,bhat05,bhat2002,bhat2004,bhat96,bec98,beck96,barrow,farrar96,shapi79,shapi821,shapi82,shiva,dunn,adey07}. The origin of these fragments extends from quasi-elastic (QE)/
projectile breakup \cite{carlin,pada}, deep-inelastic (DI) transfer
and orbiting \cite{bhat05,bhat2004,shapi79,shapi821,shapi82,shiva,dunn},
to fusion-fission (FF)
\cite{moretto,Sanders91,Matsuse97,dha2,Szanto97,Szanto96} processes;
and in some cases the structure of the nuclei has been found to play
an important role. In most of the reactions studied, the observed
fully energy-damped yields of the fragments have been successfully
explained in terms of fusion-fission (FF) mechanism
\cite{moretto,Sanders91,Matsuse97,dha2,Szanto97,Szanto96}. However,
the reactions involving
$\alpha$ - cluster nuclei (e.g., $^{20}$Ne + $^{12}$C \cite{shapi79,shapi821}, $^{24}$Mg + $^{12}$C \cite{dunn}, $^{28}$Si + $^{12}$C \cite{shapi82,shapi84} etc.) deserved special attention, where the observations of large enhancement in yield and/or resonance-like
excitation function in a few outgoing channels have been indicative of a competitive role played by the deep-inelastic orbiting mechanism
\cite{shapi79,shapi821,shapi82,shiva}. In the FF mechanism, a completely
equilibrated compound nucleus (CN) is formed, which decays into
various exit channels. The decay probability is governed by the
available phase space and barrier penetration probabilities for the
respective decay channels. The process occurs in a similar time
scale which is required for the complete relaxation of the entrance
channel energy and angular momentum. On the other hand, deep
inelastic orbiting may be described in terms of the formation of a
long-lived, dinuclear molecular complex \cite{shiva}, which acts as
a ``door way to fusion", with a strong memory of the entrance
channel. In this picture, the interacting ions are trapped in a more
deformed configuration than that of the compound nucleus (trapped in
the pocket of the ion-ion interaction potential due to combined
effects of Coulomb and centrifugal barriers). Both orbiting and
fusion-fission processes occur on similar time scale. In addition to
that, for the light heavy-ion systems, the shapes of the orbiting
dinuclear complexes are quite similar to the saddle and scission
shapes obtained in course of evolution of the FF process. Therefore
it is difficult to differentiate the signatures of the two
processes.

The enhancement of fully energy damped reaction
yields in light systems was first observed in the study of $^{20}$Ne + $^{12}$C inelastic
scattering at backward angles
\cite{shapi79}, where large cross sections have been
observed in inelastic scattering yields near 180$^{o}$. Subsequently, orbiting was observed in $^{28}$Si + $^{12}$C \cite{shapi82,shapi84} and $^{24}$Mg + $^{12}$C \cite{dunn} reactions. Detailed
study of $^{28}$Si + $^{12}$C system revealed that, at lower bombarding
energies, the excitation spectra for the $^{12}$C fragments were
dominated by single excitation and mutual excitations of the
$^{12}$C and $^{28}$Si fragments, whereas at higher bombarding
energies, the dominant strength for all these channels shifted to
higher excitation energies \cite{shapi84}. For the higher bombarding energies, 
the most probable Q-values were found to be independent of
detection angles and the resulting angular distributions were found
to have d$\sigma$/d$\Omega \propto 1/\sin{\theta_{c.m.}}$ like
angular dependence --- characteristic of a long-lived, orbiting,
dinuclear complex. Similar results have been obtained for $^{20}$Ne
+ $^{12}$C system \cite{shapi79,shapi821}, where resonance-like
behaviour was also found in the excitation functions for several
outgoing channels, which was similar to the observation made for symmetric
$^{16}$O + $^{16}$O system \cite{gobbi73,kovar,kolata,tserr78}. Enhancements of large
angle, binary reaction yields have also been observed in somewhat heavier
$^{28}$Si + $^{28}$Si, $^{24}$Mg + $^{24}$Mg systems
\cite{Sanders99}, where significant non resonant background yield
was observed at higher excitation energies. The general pattern which unfolds from these studies clearly suggests that the enhancements are manifestations of dynamics of damped nuclear reactions
involving a large number of channels, rather than due to specific
structure effect appearing only in a few select channels.

The enhancement in elastic and inelastic channels may
be explained in terms of a long lived dinuclear configuration that
decays back to entrance channel due to weak absorption which
inhibits the orbiting configuration from spreading into compound
nuclear states. However, the enhancement in the elastic
channel can be explained with the assumption of weak absorption of grazing partial
waves only; on the contrary, deep-inelastic orbiting phenomenon in general suggests
weak absorption in the angular momentum window between the critical
angular momentum of fusion, $l_{cr}$, and the reaction grazing angular
momentum, $l_{gr}$. Besides, substantial mass and
charge transfer, due to the rapid mass equilibration
in light systems, would also occur during the evolution of the
orbiting dinuclear complex. So, the rearrangement channels are
also of interest in probing the dynamics of the orbiting process
involving light nuclear systems.

It is, therefore evident that, though some qualitative understanding about the phenomenon of deep-inelastic orbiting reaction, in general ($i.e.$, correlation with number of open reaction
channels \cite{beck94}, or, alternatively, to weak absorption) has been arrived at, precise mechanism of the process is
still unknown. The deep-inelastic orbiting process has been observed in several light $\alpha$-like systems, for example, $^{20}$Ne + $^{12}$C \cite{shapi79,shapi821}, $^{24}$Mg + $^{12}$C \cite{dunn}, $^{28}$Si +
$^{12}$C \cite{shapi82,shapi84}, $^{28}$Si + $^{16}$O \cite{oliv}
systems, where the number of open channels are small ($\sim$ 10) \cite{beck94}. However, $^{16}$O (116 MeV) + $^{28}$Si reaction \cite{bhat2004} showed different behaviour so far as the shape of the energy
distributions, variation of $<Q>$ values with angle, and yields of
the fragments are concerned. For a better understanding of the orbiting process,
it is interesting to study how the orbiting process evolves
with energy. Intuitively, survival of long lived dinuclear
configuration  other than fused composite is less probable at higher
excitations and there are also indications that entrance channel
effect becomes smaller at higher energies \cite{Sanders99}. Shapira {\it et al.} \cite{shapi821} made detailed study of $^{20}$Ne  + $^{12}$C
system in the energy range E$_{lab}$  =  54 -- 81  MeV and showed
that there was large enhancement of strongly damped yields, the
characteristic of a long-lived orbiting $^{20}$Ne + $^{12}$C
dinuclear system. The aim of the present paper was to extend the investigation on fragment yield from $^{20}$Ne + $^{12}$C reaction at higher excitation energies, which
might allow us to have a better understanding of orbiting vis-$\grave{a}$-vis
fusion-fission processes for $^{20}$Ne + $^{12}$C system. If a long-lived rotating dinuclear complex is formed in these reactions, mass and charge transfer should also occur, which leads to typical deep-inelastic reaction yields. Therefore, backangle measurements for rearrangement channels became of interest. In addition to that the excitation function measurements, for the dependence of the average total kinetic energy ($E_{K}^{tot}$) loss on bombarding energy for each binary exit channel, would provide an important probe of the dynamical properties of the long-lived dinuclear complex. With this motivation, we have studied the
fragment emission spectra from the reaction $^{20}$Ne  + $^{12}$C at
E$_{\rm lab}$ =  145, 158, 170, 180 and  200  MeV, respectively. A part of the present data has already been reported \cite{bhat05}, which showed enhancement in yield of Carbon and Boron fragments well above the standard statistical model predictions.

The paper has been arranged as follows. Section II describes the
experimental procedures. In Sec. III we present the analysis of
$^{20}$Ne + $^{12}$C data. The results have been discussed in Sec
IV. Finally, the summary and conclusions are presented in Sec. V.

\section{Experiments}
The  experiment was performed using  accelerated $^{20}$Ne ion beams of
energies 145, 158, 170, 180 and 200 MeV, respectively, from the
Variable Energy Cyclotron at Kolkata. The target used was $\sim$ 550
$\mu$g/cm$^2$  self-supporting  $^{12}$C.  Different fragments (5
$\leq$ Z $\leq$ 13) have been  detected using two solid state
[Si(SB)] telescopes ($\sim$ 10 $\mu$m $\Delta$E, 300 $\mu$m E)
mounted on one arm  of the  91.5  cm scattering  chamber. Two solid state telescopes ($\sim$ 50 $\mu$m, 100 $\mu$m $\Delta$E [Si(SB)]
and 5 mm E [Si(Li)])  were mounted on the other arm  of  the
scattering chamber for the detection of light charged particles and
light fragments (1 $\leq$ Z $\leq$ 4); the same detectors were also
used as monitor detectors for normalization purposes. Typical solid
angle subtended by each detector was $\sim$0.3 msr. The telescopes
were calibrated using elastically scattered $^{20}$Ne ion from Au,
Al targets and a Th-$\alpha$ source.  The systematic  errors in  the
data, arising from  the uncertainties in the measurements of solid
angle, target thickness and the calibration of current digitizer
have been estimated  to be $\approx$  15\%. Part of these
uncertainties are due to the extrapolation procedures employed to
extract the fully damped yields from the sequential decay components
by the use of Monte Carlo simulations described in Sec. IV A.

\section{Results}
\subsection{Energy distribution} 
\begin{figure} 
{\epsfig{file=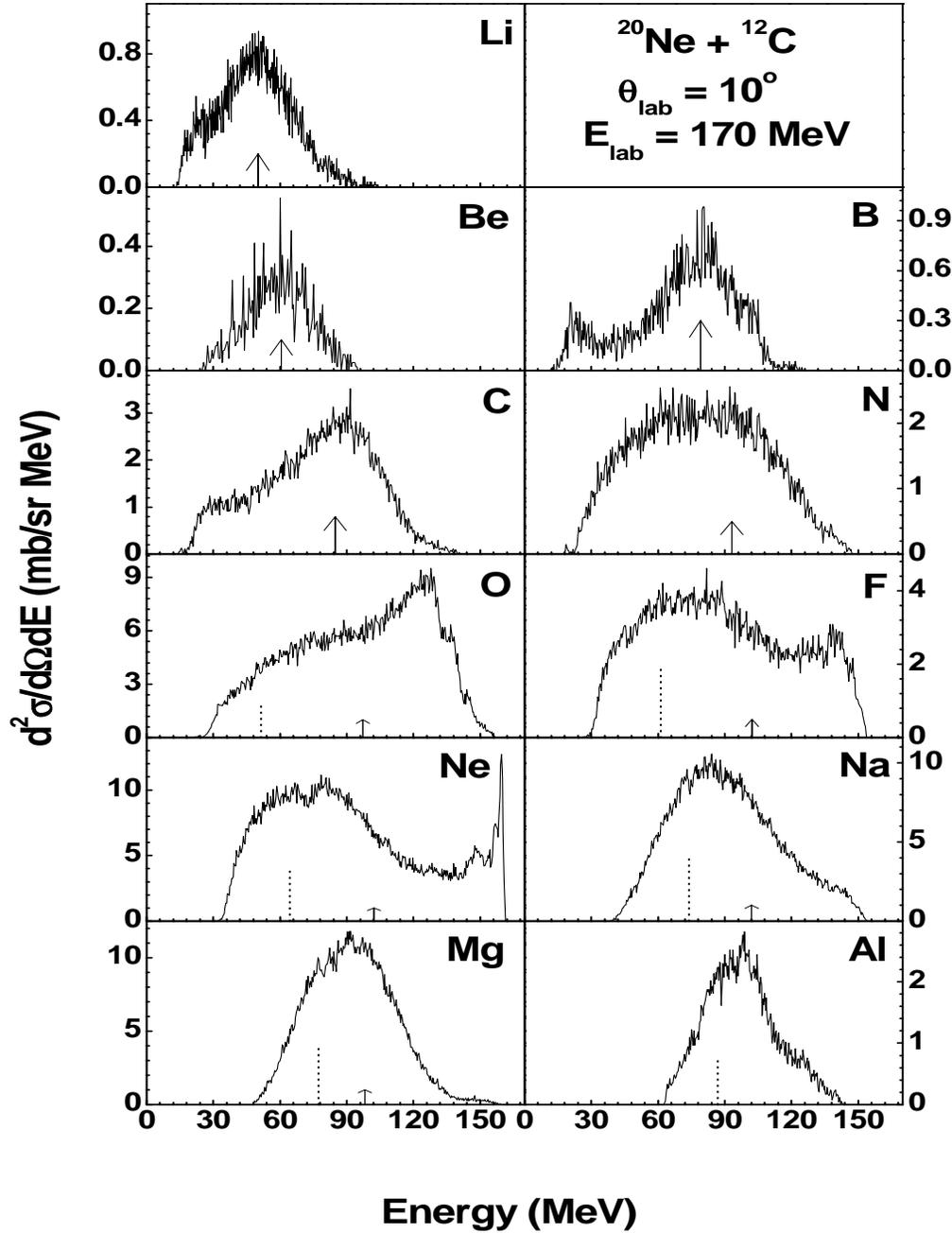,width=14.0cm,height=18.0cm.}}
\caption{ Inclusive energy distributions of different fragments
emitted in the reaction $^{20}$Ne (170 MeV) +$^{12}$C at $\theta_{lab}$ =
10$^{o}$. The arrow corresponds to the
expected fission fragment kinetic energy. The dashed line indicates
the average energy of the recoiling nucleus.} \label{nec1}
\end{figure}

Inclusive  energy  distributions
for  various fragments (3 $\leq$ Z $\leq$ 13) have been measured in
the angular range 10$^{o}$ - 50$^{o}$ for all the bombarding
energies. This covered backward angles in the center-of-mass (c.m.)
frame, because of the inverse kinematics of the reactions. Typical energy spectra of the emitted fragments (3 $\leq$ Z $\leq$ 13) obtained at an angle 10$^{o}$ at $E_{lab}$ = 170 MeV are shown in Fig.~\ref{nec1}. It is evident that the energy spectra for lighter fragents (3 $\leq$ Z $\leq$ 6) exhibit strong peaking in energy. The peaks are nearly Gaussian in shape having its centroid at the expected kinetic energies for the fission fragments obtained from the Viola systematics corrected by the corresponding asymmetry factors \cite{viola,beck92} and are shown by arrows in Fig.~\ref{nec1}. The shapes of the energy spectra for the other heavier fragments (7 $\leq$ Z $\leq$ 13) are quite different from those obtained for the lighter ones. The additional contributions from DI and QE processes have been seen in the higher energy part of the spectra. Moreover, there are contributions from the recoiling nuclei (energy corresponding to $v_{CN} \cos \theta_{lab}$, shown by dashed lines in Fig.~\ref{nec1}). All these contibutions, other than fission fragments, fall off rapidly as one moves away from the grazing angle. In this paper, we report the results from the lighter fragments (Z = 3--7). 

\begin{figure} 
{\epsfig{file=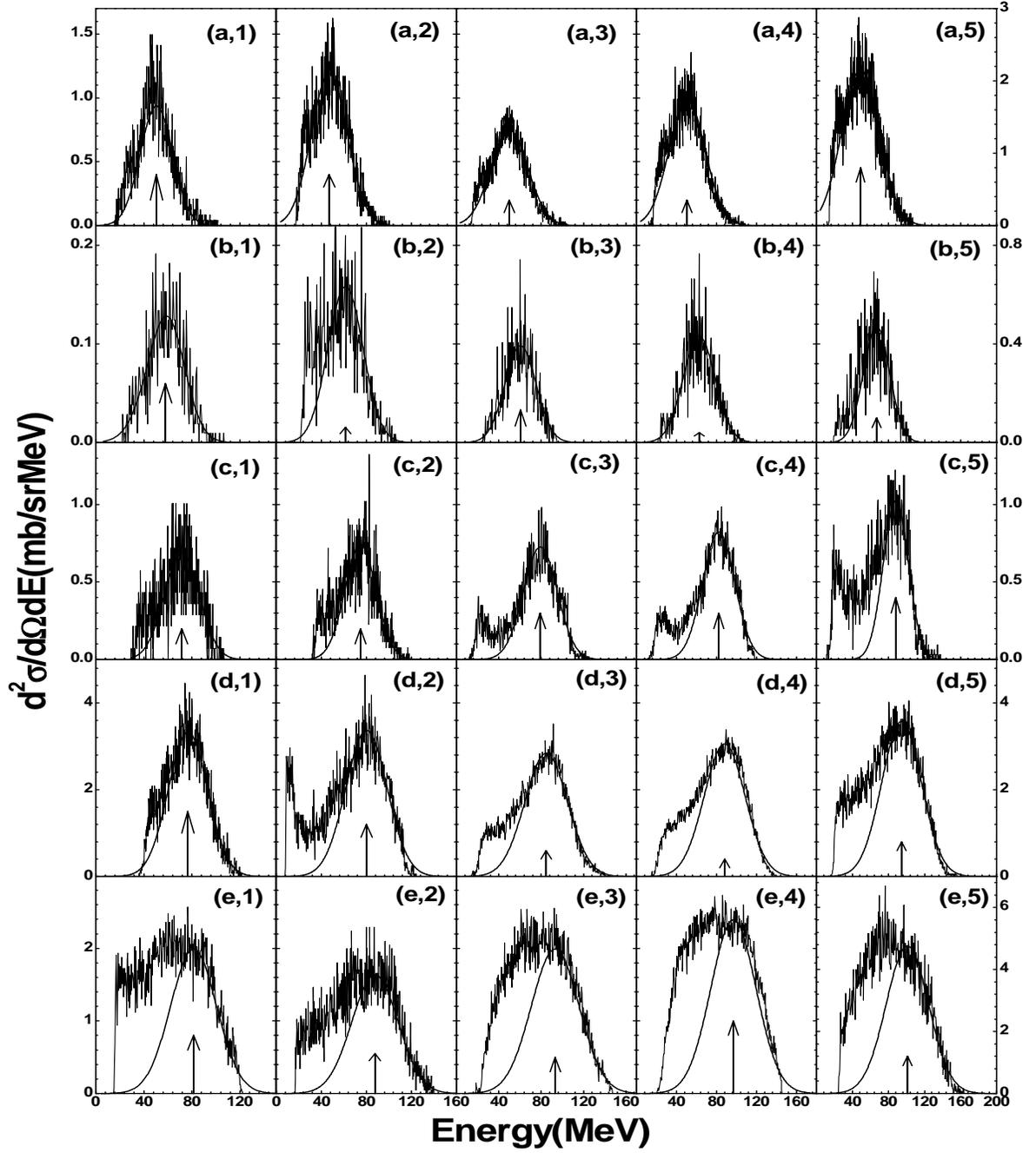,width=16.0cm,height=18.0cm.}}
\caption{ Inclusive energy distributions for the fragments Lithium (a), Beryllium (b), Boron (c), Carbon (d) and Nitrogen (e) emitted in the reaction $^{20}$Ne +$^{12}$C at an angle 10$^{o}$ for bombarding energies 145 MeV (1), 158 MeV (2), 170 MeV (3), 180 MeV (4) and 200 MeV (5), respectively. The arrow corresponds to the
centroid of the fitted Gaussian distribution (solid curve).} \label{nec2}
\end{figure}

The inclusive energy distributions for the fragments Lithium (Z =
3), Beryllium (Z = 4), Boron (Z = 5), Carbon (Z = 6) and Nitrogen (Z
= 7) obtained at an angle 10$^{o}$ at various bombarding energies
are shown in Fig.~\ref{nec2}. It is observed that at all bombarding energies the energy spectra of the ejectiles (Li, Be, B, C, N) are nearly Gaussian in shape and they have been fitted with a single Gaussian. The non-Gaussian shapes at the low-energy side of
the spectra correspond to sequential decay processes -- which can be
simulated by Monte-Carlo statistical model calculations (described
in Sec. IV A). The Gaussian fits
so obtained are shown by solid lines in Fig.~\ref{nec2}. The centroids (shown by
arrows) are found to correspond to the scission of deformed
dinuclear configuration \cite{shiva,viola,beck92}. This
suggests that, in all cases, the fragments are emitted from fully
energy relaxed composite -- as expected for both FF and orbiting
processes. The increasing yields at lower energies may also be due
to the second kinematical solution, which is a signature of binary
nature of emission process.

\subsection{Average velocity}
\begin{figure} 
{\epsfig{file=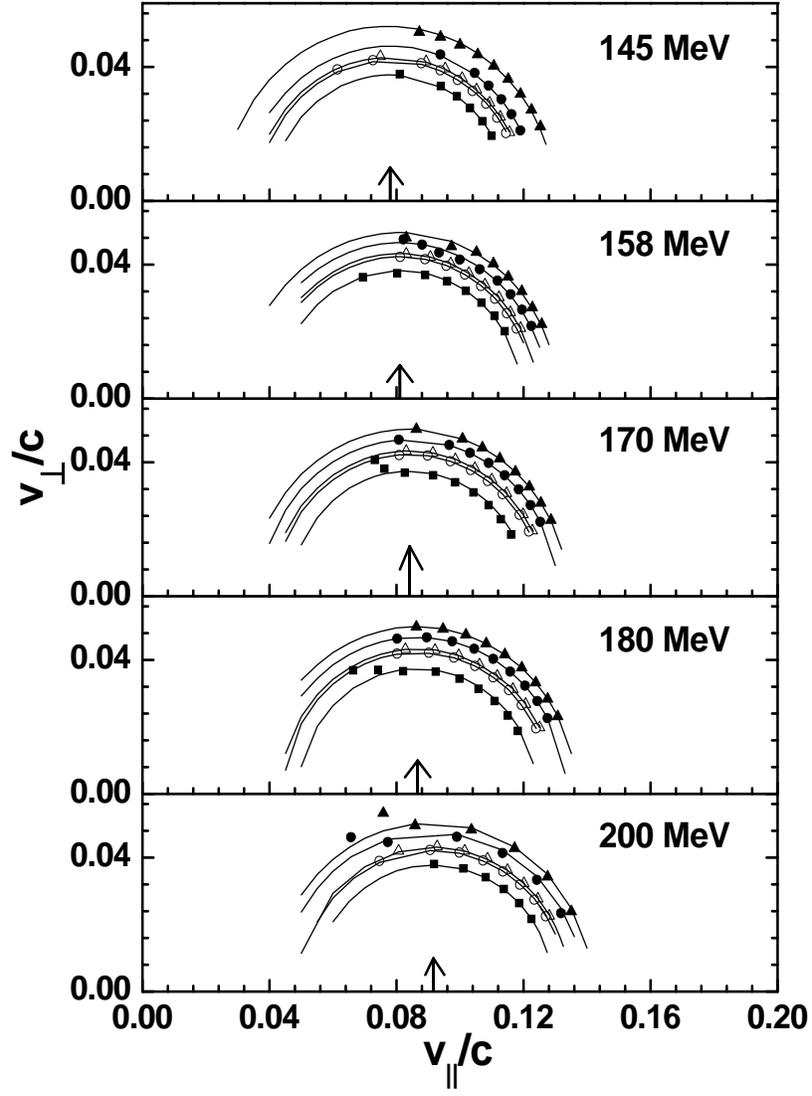,width=11.0cm,height=15.0cm}}
\caption{The average velocities of the fragments plotted in
$v_\parallel$ vs $v_\perp$ plane at different bombarding energies.
The average velocities are denoted by filled triangles (Li), filled
circles (Be), open triangles (B), open circles (C) and filled
squares (N). The arrows correspond to the compound nucleus
velocities.} \label{nec3}
\end{figure}

The average velocities of the fragments have been computed from the
measured energies and from the Z values using the empirical relation
proposed by Charity {\it et al.} \cite{charity}:
\begin{equation}
A = Z \times (2.08 + 0.0029 \times Z).
\end{equation}
The average velocities of the fragments obtained at different bombarding energies have been plotted in the
$v_\parallel$ vs. $v_\perp$ plane in Fig.~\ref{nec3}. It is seen
that at all energies, the average velocities of different fragments fall on a circle centered around the respective
$v_{CN}$, the compound nuclear velocity. This suggests that at all bombarding energies the average
velocities (as well as kinetic energies) of the fragments are
independent of the c.m. emission angles and indicates that at all these energies the fragments are emitted
from a fully equilibrated CN-like source with full
momentum transfer. The magnitude of the average fragment velocities
({\it i.e.}, the radii of the circles in Fig.~\ref{nec3}) decreases
with the increase of fragment mass, which is indicative of the
binary nature of the emission.

\subsection{Angular distribution}
\begin{figure} 
{\epsfig{file=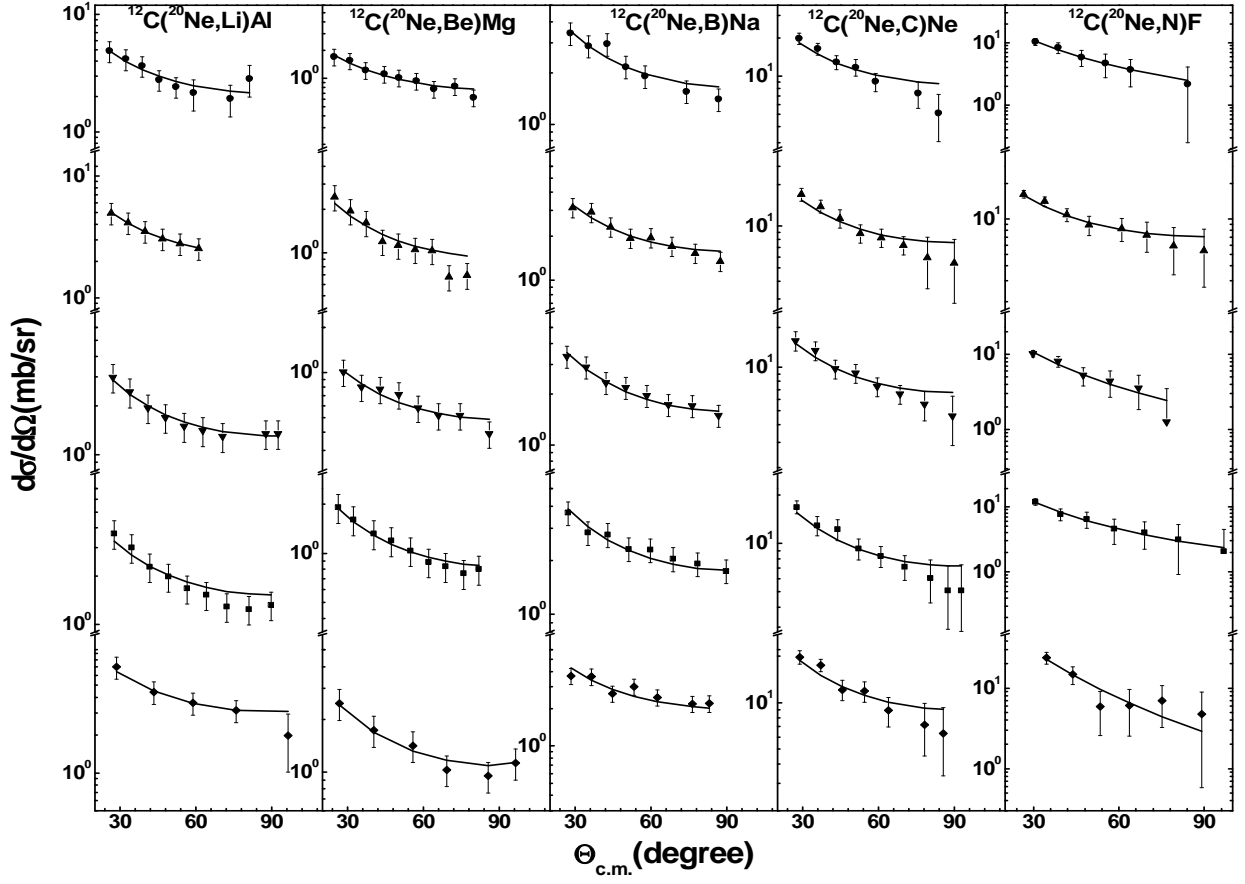,width=17.0cm,height=12.0cm}}
\caption{The c.m. angular distributions of fragments (Z = 3 -- 7) obtained at different bombarding energies. Solid circles (145 MeV), triangles (158 MeV), inverted triangles (170 MeV), squares (180 MeV) and diamonds (200 MeV) correspond to the experimental data and the solid lines are $f(\theta_{c.m.}) \sim 1/\sin\theta_{c.m.}$ fit to the data.}
\label{nec4}
\end{figure}

The center-of-mass (c.m.) angular distributions of the fragments
(Li, Be, B, C, and N) emitted in the $^{20}$Ne (145, 158, 170, 180
and 200 MeV) + $^{12}$C reactions are shown in Fig.~\ref{nec4}. The transformations from the
laboratory to center-of-mass system have been done with the
assumption of a two body kinematics averaged over total kinetic
energy distributions. The c.m. angular distributions
of these fragments obtained at all bombarding energies follow the
1/sin$\theta_{c.m.}$ -like variation (shown by solid lines in Fig.~\ref{nec4}) --- which further corroborate the conjecture of emission
from fully equilibrated composite.

\subsection{Average kinetic energy}
\begin{figure}
{\epsfig{file=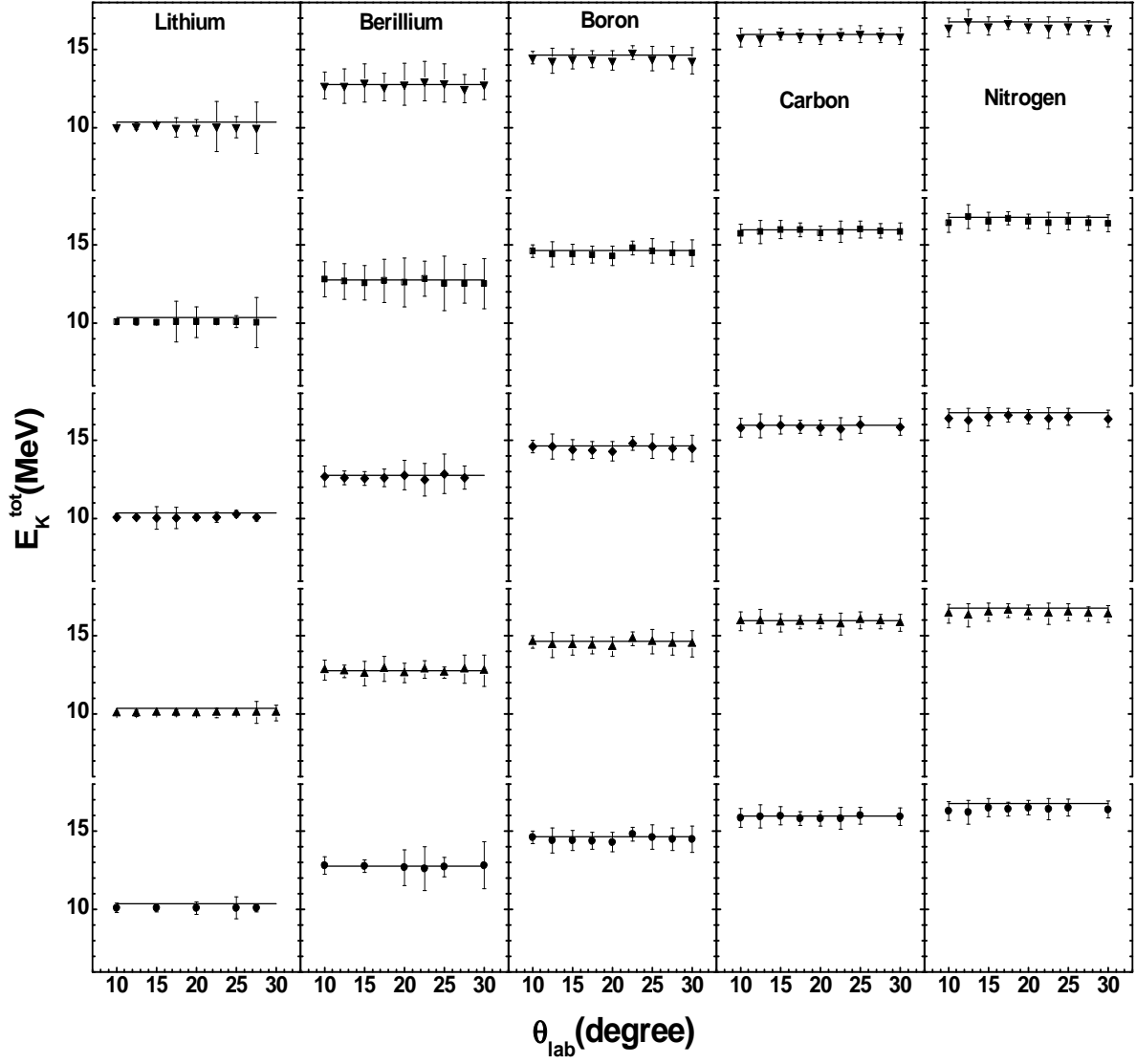,width=16.0cm,height=15.0cm}}
\caption{ Average kinetic energy of different fragments obtained at
E$_{lab}$ = 145, 158, 170, 180 and 200 MeV (denoted by inverted
triangle, square, diamond, triangle and circle, respectively)
plotted as a function of laboratory angle.} \label{nec5}
\end{figure}

The average total kinetic energies in the center-of-mass,
$E_{K}^{tot}$, for the fragments ($3 \leq Z \leq 7$) obtained at all bombarding energies, have been displayed as a function of scattering angle in Fig.~\ref{nec5}. The
average fragment kinetic energies in the center-of-mass have been
obtained from the respective laboratory values assuming two body
kinematics. It is observed from Fig.~\ref{nec5} that $E_{K}^{tot}$
values are almost constant for each of the exit channel. The near constancy of $E_{K}^{tot}$ indicates
that at all energies the lifetime of the dinuclear complex is longer than the time
needed to completely damp the energy in the relative motion
\cite{egger,braun,cor,nato}. The predictions of Viola systematics
\cite{viola} for fission fragment kinetic energies, corrected by an
asymmetric factor \cite{beck92}, have been shown by solid lines in
Fig.~\ref{nec5}. The $E_{K}^{tot}$ values predicted from Viola
systematics are found to be in good agreement with the experimental
data at all bombarding energies. 

\subsection{Average Q-value distribution}
\begin{figure}
{\epsfig{file=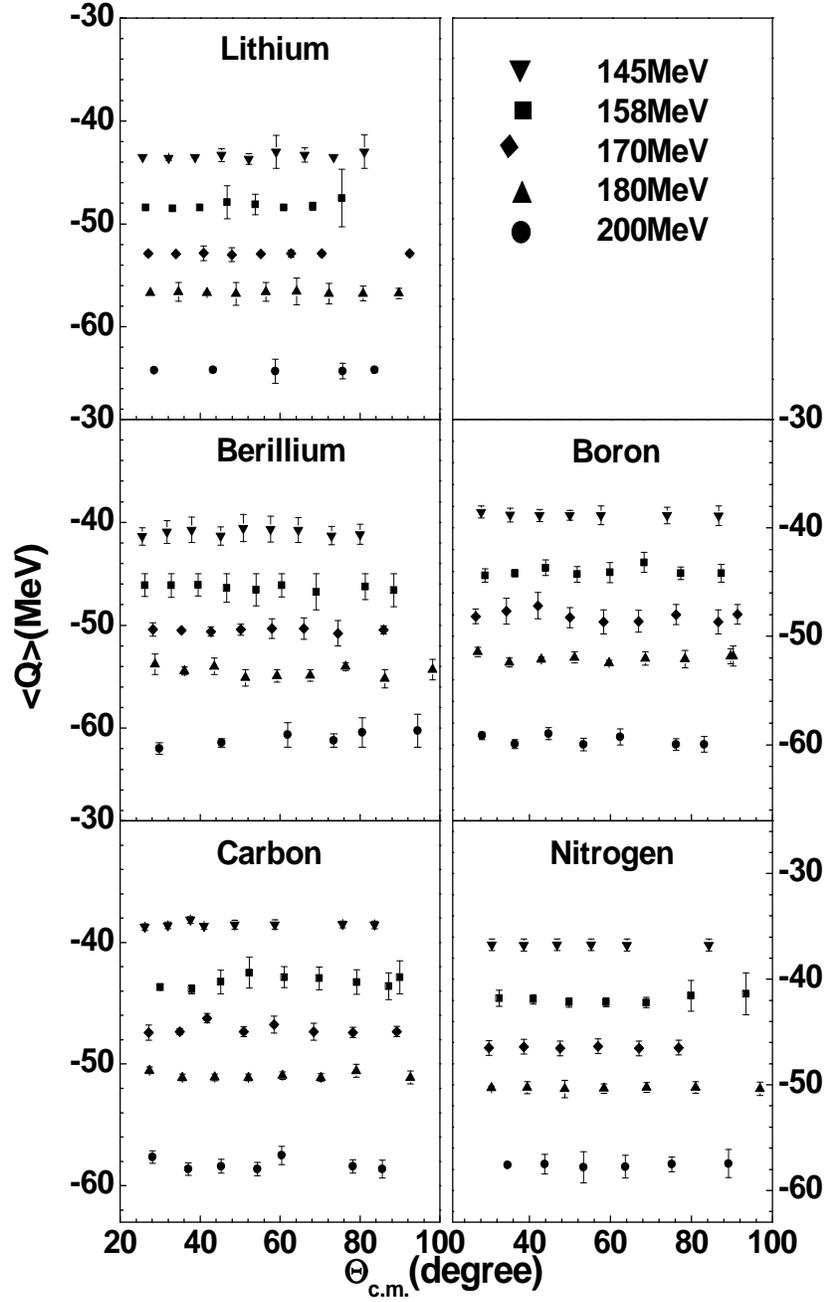,width=12.0cm,height=18.0cm}}
\caption{ Average Q-values of different fragments obtained at
E$_{lab}$ = 145, 158, 170, 180 and 200 MeV (denoted by inverted
triangle, square, diamond, triangle and circle, respectively)
plotted as a function of c.m. emission angle.}
\label{nec6}
\end{figure}

The variations of average Q-value, $<Q>$, with center-of-mass
emission angle for the fragments Li, Be, B, C, and N obtained at
different bombarding energies are shown in Fig.~\ref{nec6}. It is
observed that the $<Q>$ of different fragments are independent of the center-of-mass emission angles at all bombarding energies. This is in contrast to the observation made earlier for other light systems ($^{16}$O (116 MeV) + $^{27}$Al, $^{28}$Si, $^{20}$Ne (145 MeV) +
$^{27}$Al,  $^{59}$Co systems), where sharp fall off of $<Q>$ with angle have been seen \cite{bhat2002,bhat2004}. The  $<Q>$ values remain nearly constant which
further suggest that at all angles, the fragments are emitted from
completely equilibrated source at all incident energies considered
here.

\subsection{Equilibrium cross-section}
The energy distributions, velocity diagrams, angular distributions and $<Q>$ distributions indicate that the yield of these fragments (Z = 3 -- 7) originates from fully energy relaxed events
associated with the decay of either compound nucleus or long-lived,
orbiting dinuclear system. A detailed investigation have been made to decipher the role played by aforementioned processes in the fragment yield by comparing the experimental yields with the theoretical
predictions of the standard statistical model \cite{pul}, extended
Hauser-Feshbach model (EHFM) \cite{Matsuse97}. The experimental angle integrated
yields of the fragments emitted in the $^{20}$Ne + $^{12}$C reaction at different bombarding energies are shown in Fig.~\ref{nec7} by solid circles (taken from Ref.~\cite{shapi821}) and
triangles. The theoretical predictions of the statistical model code CASCADE \cite{pul} are shown by solid lines in Fig.~\ref{nec7}. The
calculations are done considering $l$ values up to the critical
angular momentum of fusion, $l_{cr}$, at each energy (given in Table \ref{tbl1}). It is observed that at all energies the
experimental yields of the fragments Li and Be are in fair agreement with the theoretical CASCADE prediction. However, the yields of the fragments B, C and N (near entrance channel) obtained at different energies are much higher than those predicted by the statistical model code CASCADE. A similar observation has been reported by
Shapira {\it et al.} \cite{shapi79,shapi821} for the same system at
lower energies for Carbon fragment. The theoretical predictions using EHFM have also been shown by dotted lines in Fig.~\ref{nec7} for the fragments B, C and N. The EHFM predictions also similar to those obtained from CASCADE calculations and the experimental yields are in fair excess of the theoretical estimates of both CASCADE and EHFM.

\subsection{Excitation energy dependence of $<Q>$}
\begin{figure}
{\epsfig{file=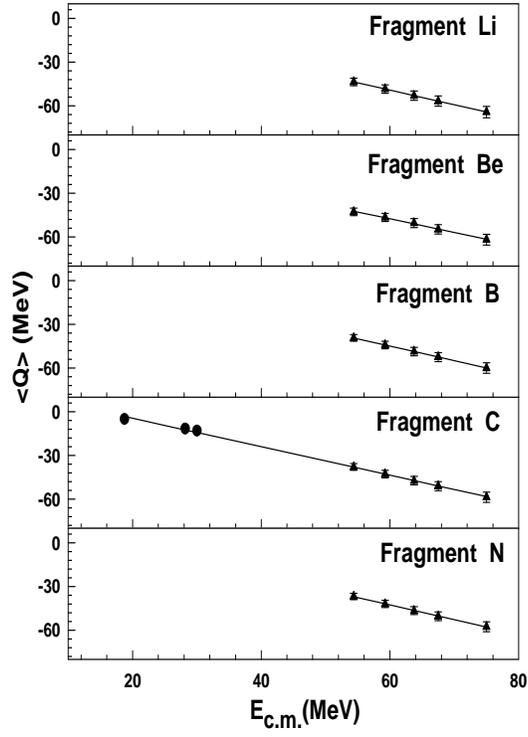,width=7.0cm,height=10cm}}
\caption{Bombarding energy dependence of the average Q-values.
The solid line shows the linear dependence of $<Q>$ with
bombarding energy. The $<Q>$ values at lower energies for Carbon
are taken from \protect\cite{shapi821}.}
\label{nec8}
\end{figure}

The average Q values ($<Q>$) for the fragments Li, Be,
B, C, and N have been plotted in Fig.~\ref{nec8} as a function of the incident energy. The linear dependence of $<Q>$ with energy provides
strong evidence that the long life time may be associated with an
orbiting phenomenon. This linear dependence of $<Q>$ can be
expressed by simple equation of the form $<Q> = c - m * E_{c.m.}$, where $m$ is the slope and $c$ is the intercept (for example, {$<Q>$ = (14.9
$\pm$ 1.0) - (0.97 $\pm$ 0.02) E$_{c.m.}$} for the fragment Carbon). The experimentally determined intercepts are found to be in fair agreement with kinetic energies calculated using Viola systematics. It is interesting to note that the $<Q>$ values obtained in the
present experiment between 145 MeV and 200 MeV fall on the same
straight line extrapolated  from the lower energy ($\sim$54 -- 81
MeV) data \cite{shapi821}. This means that the energy relaxation is
complete for the fragment emission studied here up to the incident
energy of 200 MeV. Moreover, it also means that the final kinetic
energy {(E$^f_{kin}$ = $<Q>$ + E$_{c.m.}$)} is nearly independent of
bombarding energy --- which may be due to the limitation on the
maximum value of angular momentum beyond which the formation of
di-nucleus is not allowed due to centrifugal repulsion
\cite{shapi84}.

\section{Discussions}
In general, the energy distributions, the angular distributions and
the total fragment yields measured for $^{20}$Ne + $^{12}$C reaction
at incident energies between 145 MeV and 200 MeV are similar to
those obtained at lower incident energies ($\sim$ 50 - 80 MeV) for
the same system (see Refs.~\cite{shapi79,shapi821}). Large energy
damping, 1/sin$\theta_{c.m.}$ dependence of angular distribution and
near constancy of $<Q>$ over a wide angular range signify that the
fragment decay originates from a long lived, fully energy
equilibrated system. However, the large enhancement of fragment
emission cross section (5 $\leq$ Z $\leq$ 7) over the statistical-model predictions leads
to the conjecture that the orbiting mechanism may still play a major
role at these energies. The possibility of these enhancements, either due to feeding from the secondary deexcitation of the heavier fragments or due to orbiting mechanism, are investigated in great details and described in the following subsections.

\subsection{Contribution from secondary decay of heavier fragments}
\begin{figure}
{\epsfig{file=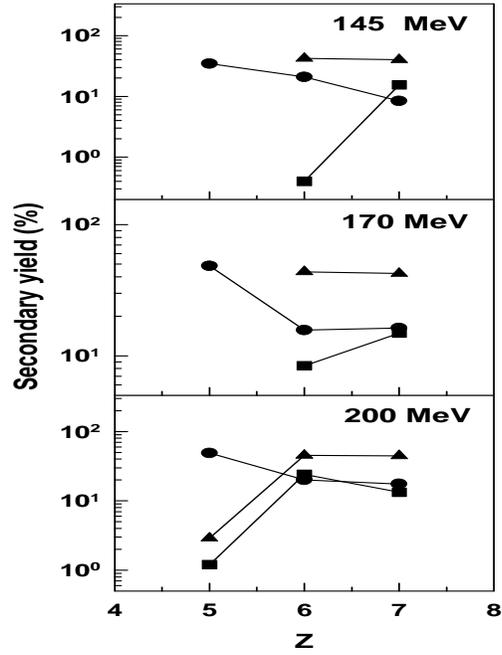,width=7.0cm,height=9.0cm}}
\caption{Percentage secondary decay contribution from primary Ne$^{*}$ (circle), F$^{*}$ (square) and O$^{*}$ (triangle) to Z = 5 -- 7.}
\label{nec9}
\end{figure}

There is a possibility that the primary heavier fragments (formed due to the binary decay of composite system) may have sufficient excitation energy to deexcite through the emission of light particles and $\gamma$-rays and contribute to the yield of lighter fragments. This additional contribution from the secondary decay increases the total elemental yield of the lighter fragments. To check whether the enhancement in B, C, and N yield could be due to
feeding from the secondary decay of heavier fragments of various
possible binary breakup combinations, we have performed detailed
simulations of secondary decay using the Monte-carlo binary decay
version of the statistical decay code LILITA \cite{lilita} and the
statistical model code CASCADE \cite{pul} and PACE4 \cite{pace2}. Secondary decay of Si$^*$
(binary channel $^{28}$Si + $^{4}$He), Al$^*$ (binary channel
$^{26,25}$Al + $^{6,7}$Li), Mg$^*$ (binary channel $^{24,25,23}$Mg + $^{8,7,9}$Be),
Na$^*$ (binary channel $^{22,21}$Na + $^{10,11}$B), Ne$^*$ (binary channel
$^{20}$Ne + $^{12}$C), F$^*$ (binary channel $^{18}$F + $^{14}$N), O$^*$ (binary channel $^{16}$O + $^{16}$O), N$^{*}$ (binary channel $^{14}$N + $^{18}$F), C$^{*}$ (binary channel $^{12}$C + $^{20}$Ne), B$^{*}$ (binary channel $^{10,11}$B + $^{22,21}$Na) and Be$^{*}$ (binary channel $^{8,7,9}$Be + $^{24,25,23}$Mg) have been studied.
Indeed the LILITA calculations (using the parameter set proposed in
the Appendix of Ref. \cite{shapi821} and assuming the excitation
energy division follows the mass ratio \cite{Sanders99}) are in
qualitative agreement with the experimental results obtained at 9
MeV/nucleon by Rae et al. \cite{Rae} for the sequential decay of
$^{20}$Ne + $^{12}$C. It was found that even at the highest
excitation energy, secondary decay of Si$^*$ and Al$^*$ do not reach up to N; the contribution of primary Mg$^{*}$, Na$^{*}$ decay to Z $\leq$ 7 were estimated to be $\lesssim$ 1\% of the primary yield and that of Ne$^*$
decay to N, C, B yield were estimated to be $\sim$ 10--20\%, $\sim$
15--20\% and $\sim$ 30--50\% of the primary yield, respectively.
Nearly $\sim$ 40--45\% of the primary O$^*$ produced through binary
exit channel $^{16}$O + $^{16}$O decays to C. The secondary decay
yields from the primary excited fragments are shown in Fig.~\ref{nec9} for different bombarding energies.
As the binary yield of O, F are small ($\sim$ 10\% of
the binary Ne yield, as estimated from CASCADE \cite{pul}), overall
secondary decay contribution from O, F are smaller than that from
Ne. Moreover, the simulations of energy distributions of the
secondary decay yield of C from Ne as well as F, O using the code
LILITA show that they peak at much lower energies (typically, at
$\sim$ 45--50 MeV for Ne, $\sim$ 48--55 MeV for F and $\sim$ 55--60 MeV for O, compared to
the peak of the experimental energy distribution at $\sim$ 75--95
MeV).

\begin{figure}
{\epsfig{file=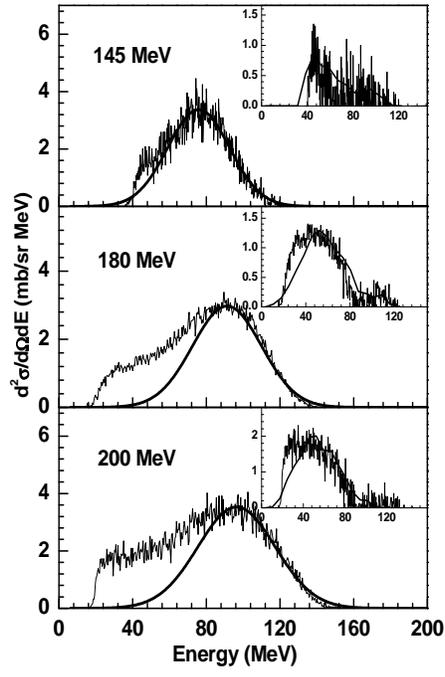,width=6.0cm,height=9.0cm}}
\caption{ Secondary decay contribution for Carbon fragments at
different energies. The energy distribution at $\theta_{lab}$ =
10$^{o}$ along with the fitted Gaussian are shown. Inset:
Distribution shows the difference spectra (total spectra -
Gaussian) and the solid line represents the total secondary decay
contribution estimated using LILITA \protect\cite{lilita}. }
\label{nec10}
\end{figure}

\begin{figure}
{\epsfig{file=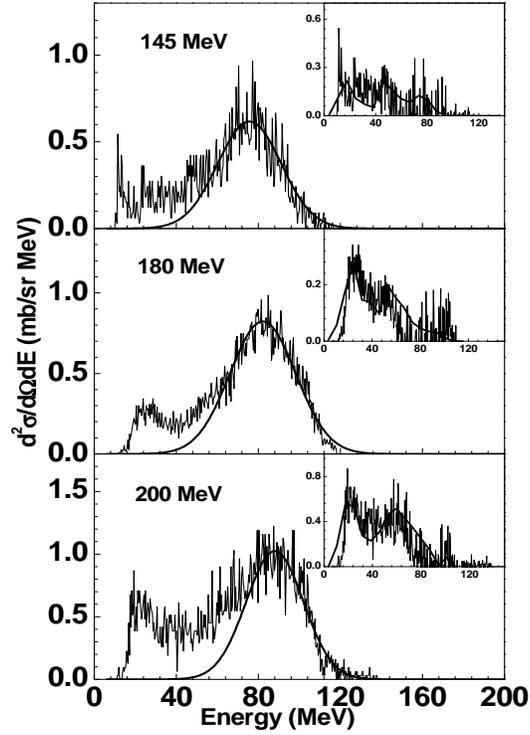,width=7.0cm,height=10.0cm}}
\caption{Same as Fig.~\ref{nec10} for Boron fragments.}
\label{nec11}
\end{figure}

\begin{figure}
{\epsfig{file=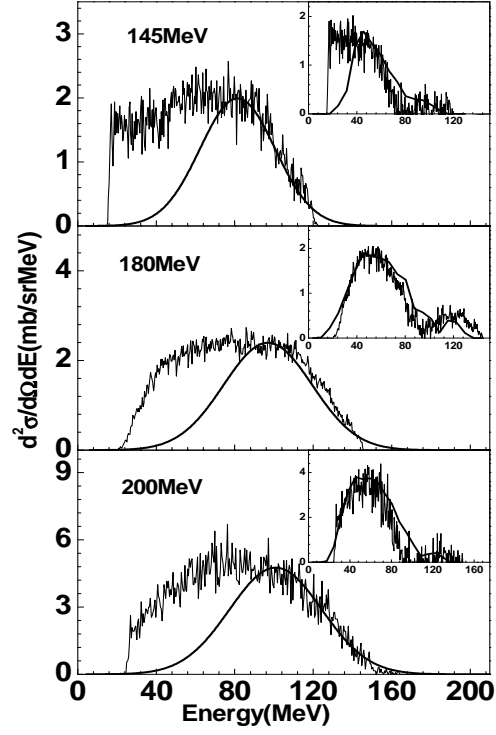,width=7.0cm,height=10.0cm}}
\caption{Same as Fig.~\ref{nec10} for Nitrogen fragments.}
\label{nec12}
\end{figure}

Now, the Gaussian fitting procedure for the extraction of primary fragment yield is fairly efficient in rejecting most of the low energy tail
(typical rejection ratio $\sim$ 25--40\% of the total yield). The
energy distributions of the secondary decay have been shown in Fig.~\ref{nec10} for Carbon, in Fig.~\ref{nec11} for Boron and in Fig.~\ref{nec12} for Nitrogen. In the inset of these figures the spectra
shown are the difference spectra; the difference between the
experimental spectra and the Gaussian fitted to the spectra. It has
been found that the secondary decay distributions reproduce the
difference spectra very well for all the cases. It is thus evident
that the secondary decay component does not interfere with the
estimated primary yield for two reasons: firstly, total secondary
decay yield is not quite large, and secondly, the Gaussian fitting
procedure for the extraction of primary yield does take care, to a
large extent, of the rejection of the contributions of the secondary
decay components as their energy distributions are different from
those of the primary components.

\subsection{Fragment cross-section}
It has been observed that the statistical model calculations do not reproduce most of the
observed experimental yields, therefore, an additional reaction component
corresponding to the orbiting mechanism has to be considered. The
large measured cross sections for B, C and N fragments led to the
suggestion that an orbiting, di-nuclear configuration is formed that
decays back to the entrance channel. After the discovery of orbiting
in the $^{28}$Si + $^{12}$C system, similar enhancements of
large-angle, binary-reaction yields are also observed in the present
data. It is expected that the orbiting mechanism will retain a
greater memory of the entrance channel than the fusion-fission
process. The trapped, dinuclear complex can either evolve with
complete amalgamation into a fully equilibrated compound nucleus or,
alternatively, escape into a binary exit channel by way of orbiting
trajectories. Orbiting can therefore be described in terms of the
formation of a long-lived di-nuclear molecular complex which acts as
a ``doorway" state to fusion with a strong memory of the entrance
channel. The equilibrium orbiting model has been used to
successfully explain both the observed cross sections and total
kinetic energy (TKE) values of the fully damped fragments for
several lighter nuclear systems at lower energies. The theoretical prediction of the equilibrium model for orbiting and fusion \cite{shiva} is denoted by dash-dotted line in Fig.~\ref{nec7} for the fragment B, C and N, and it also fails to explain the large enhancement in the fragment yield. The curve displayed in Fig.~\ref{nec7} represents the ``best fit" that can be obtained by the orbiting model with a reasonable choice of
the Bass potential parameters (strengths, short range, and long
range of the proximity potential). It is, therefore, evident that both the equilibrium orbiting
and statistical decay (CASCADE, EHFM) models result in comparable
disagreement with the data. It may be interesting to note here that
Shapira {\it et al.} studied the same reaction at lower energies
\cite{shapi79,shapi821} and came to the conclusion that the large
enhancements in the energy damped fragment yield observed at those
energies might be due to nuclear orbiting phenomenon.

The shortcomings of the equilibrium model for orbiting
does not imply that the presence of an orbiting mechanism, as
distinct from fission, can be ruled out. On the contrary, there may
be a large orbiting-like contribution from non fusion window (in the
angular momentum window $l_{cr} \leq l \leq l_{gr}$). This is
consistent, at least qualitatively, from the fact that, CASCADE
calculation \cite{pul} performed with $l$ values up to  $l_{gr}$
(shown by dashed lines in Fig.~\ref{nec7}) is found to
reproduce the data fairly well. The values of $l_{gr}$ at different bombarding energies are given in Table~\ref{tbl1}. Yields in the transfer channels (B, N,
for example) are also found to be strongly affected by the orbiting
process (yield enhancement), which may be due to stochastic nucleon
exchanges during long lifetime of the dinuclear system.

\begin{figure}
{\epsfig{file=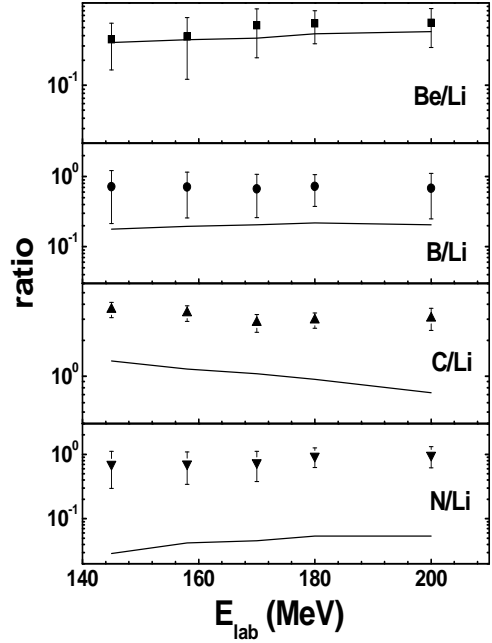,width=7.0cm,height=9.0cm}}
\caption{Bombarding energy dependence of the ratio of
angle-integrated fragment yield. The Beryllium to Lithium (square),
Boron to Lithium (circle), Carbon to Lithium (triangle), and
Nitrogen to Lithium (inverted triangle) ratios are shown. The solid
line shows the theoretical prediction using CASCADE.} \label{nec13}
\end{figure}

In Fig.~\ref{nec13}, we show the ratio of Beryllium to Lithium
(square), Boron to Lithium (circle), Carbon to Lithium (triangle)
and Nitrogen to Lithium (inverted triangle) yield as a function of
bombarding energies and the corresponding statistical model (CASCADE)
calculations (solid lines). It is found that the observed
Beryllium to Lithium ratio is well explained with the statistical
model calculations. However, the other observed ratios are higher
than the theoretically calculated ratios. This implies the dominance
of orbiting yield over the compound nucleus yield.

\subsection{Comparison of the $^{20}$Ne + $^{12}$C and $^{19}$F + $^{12}$C reactions}

The large orbiting yields that account for the largest part of the fully
damped yields of $^{20}$Ne + $^{12}$C can be qualitatively understood in the
framework of the Number of Open Channels (NOC) model \cite{beck94,beck95}.
The calculated NOC are shown in Table~\ref{tbl1} for both the $^{20}$Ne + $^{12}$C
and the $^{19}$F + $^{12}$C reactions \cite{beck94} at several excitation energies along
with the measured and calculated (fusion-fission cross sections
as predicted by CASCADE) fully damped yields. The NOC for 
$^{20}$Ne + $^{12}$C exhibits the characteristic minimum for a grazing 
angular momentum of approximately $l_{gr}$ = 30$\hbar$ \cite{beck94}. This very deep 
minimum (NOC = 4.5) explains: (i) why resonant structures have been 
observed to be significant in $^{20}$Ne + $^{12}$C \cite{kolata,tserr78} and, (ii) why the orbiting yields observed for C fragments are much larger than the 
CASCADE predictions. The comparison with $^{19}$F + $^{12}$C is instructive 
at $E^{*}$ = 60 MeV \cite{bhat96} (the corresponding value of the yield of fragment C for 
$^{20}$Ne + $^{12}$C, as given in Table~\ref{tbl1}, has been extracted from Fig.~\ref{nec7} to permit a direct comparison 
with $^{19}$F + $^{12}$C). The large NOC value for $^{19}$F + $^{12}$C \cite{beck94}
(almost order of magnitude bigger) is consistent with the fact that 
essentially no resonances have been observed in this system 
\cite{Aissaoui94,Aissaoui97}. This was confirmed by the times scale 
measurements of Suaide {\it et al.} \cite{Suaide02} who found that 
fusion-fission (with high NOC values), a very slow 
mechanism, is more competitive than a faster process such as orbiting 
(with small NOC values) in $^{19}$F + $^{12}$C. It is worth noting from 
Table~\ref{tbl1} that CASCADE predicts almost identical fusion-fission cross 
section for both reactions at $E^{*}$ = 60 MeV. On the other hand, due to the survival of 
orbiting at energies larger than 7 MeV/nucleon, the fully damped yields 
are much more than a factor of two bigger for the $^{20}$Ne + $^{12}$C system 
at $E^{*}$ = 60 MeV.

\section{Summary and Conclusions}
The inclusive double differential cross-section for fragments having
Z = 3~--~7 emitted in the reaction $^{20}$Ne ($\sim$ 7~--~10
MeV/nucleon) + $^{12}$C have been measured. Total emission
cross-section for the fragments Li to N have been estimated from the
experimental distributions. The c.m. angular distributions for the
fragments at all the bombarding energies are found to have a 1/$\sin
\theta_{c.m.}$-type of dependence which signifies the emission of
these fragments from a long-lived equilibrated composite. The
average velocity plots in $v_{\parallel}$ vs. $v_{\perp}$ plane
indicate that the fragments are emitted from fully equilibrated
source moving with compound nucleus velocity. The average kinetic
energy and the average Q-value of the fragments are independent of
the emission angles. This also suggests the emission from a
long-lived, equilibrated composite. The angle-integrated
cross-section for Li and Be fragments agree well with the
theoretical predictions of statistical model but the yield of B, C and N fragments (near to entrance channel) are in excess with the theoretically predicted values. This indicates the presence of other type of reaction process, namely orbiting. In contrast, the study of the nearby system $^{19}$F (96 MeV) + $^{12}$C clearly showed that the fragments are emitted in the fusion-fission process \cite{bhat96}. Low values of NOC \cite{beck94} obtained for $^{20}$Ne + $^{12}$C system as compared to the same obtained for $^{19}$F + $^{12}$C system also confirms the conjecture of survival of orbiting in $^{20}$Ne + $^{12}$C system at higher excitations. It is interesting to mention at this point that, $^{16}$O, $^{20}$Ne + $^{28}$Si systems \cite{bhat2004,barrow,farrar96}, even though $\alpha$-like, do not show the characteristics
of orbiting at these energies, but orbiting-like behaviour has been
observed for $^{28}$Si + $^{16}$O reaction at lower energies
\cite{Szanto97}.

The present analysis also indicates that the enhancement in fragment yield for $^{20}$Ne + $^{12}$C reactions can not be explained by the equilibrium orbiting model \cite{shiva}. This may be due to the fact that, the equilibrium orbiting model in its present form seems to be inadequate to explain the phenomena at higher excitations, and a
more complete understanding of orbiting and vis-a-vis the angular
momentum dissipation (which plays a crucial role in defining
orbiting trajectories and yield) will be required.

\vspace{0.5cm}
\noindent
{\bf Acknowledgements} \\
The authors like to thank the cyclotron operating
crew for smooth running of the machine, and H. P. Sil for the
fabrication of thin silicon detectors for the experiment. One of the
authors ( A. D. ) acknowledges with thanks the financial support
provided by the Council of Scientific and Industrial Research,
Government of India.

\begin{figure}
{\epsfig{file=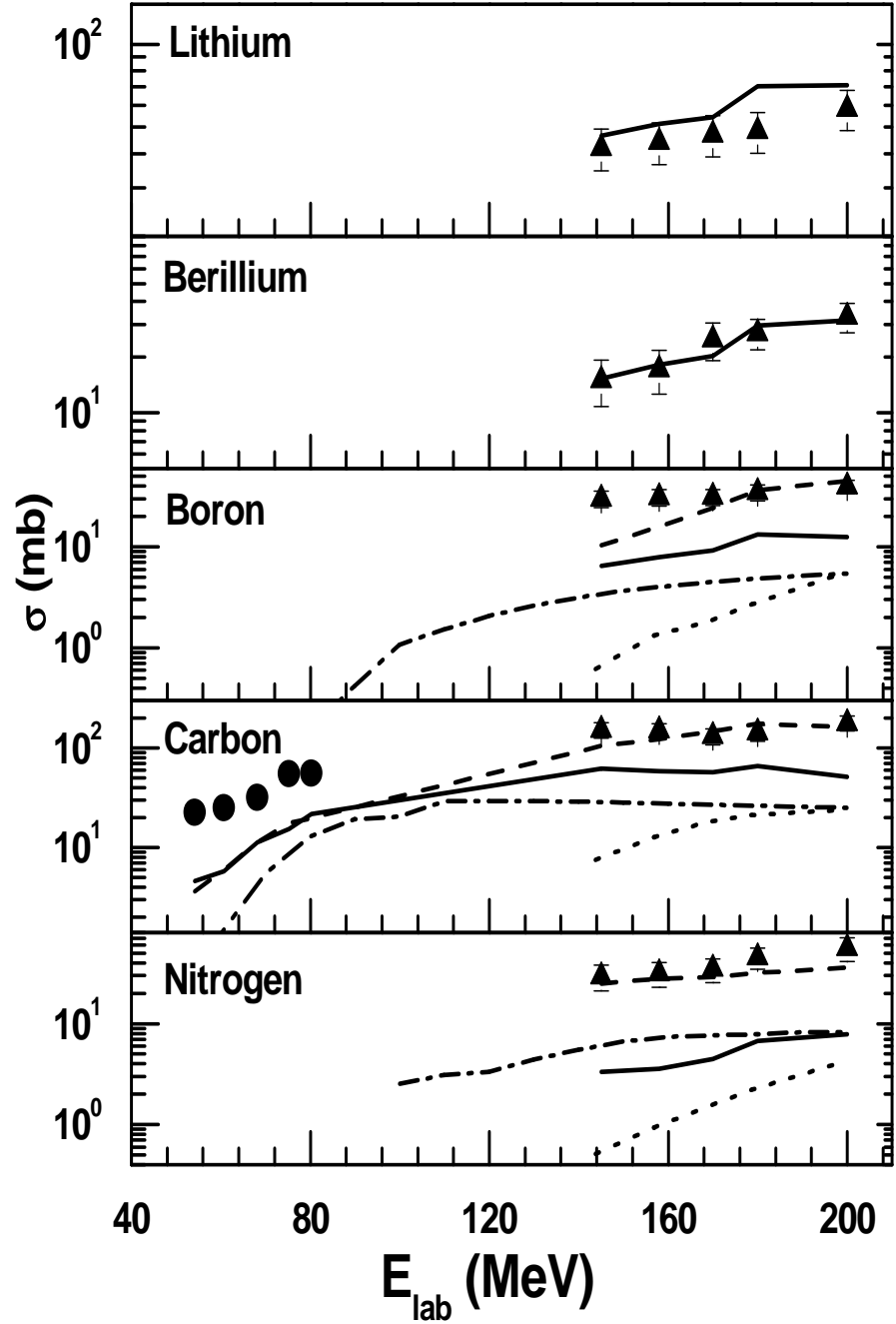,width=12.0cm,height=18.0cm}}
\caption{The excitation functions for the angle-integrated yield of
the fragments. Triangles are the present data; lower energy data
(filled circles) for Carbon fragments are taken from
\protect\cite{shapi821}. The solid curves are the predictions of the
statistical model. The dash-dotted curves for B, C and N are
prediction of equilibrium orbiting model and the dotted curves are
the same from EHFM \protect\cite{Matsuse97}. The dashed curves show CASCADE calculations using grazing angular momentum.} \label{nec7}
\end{figure}

\begin{table} [h]
\caption{The angular momentum values, NOC and C fragment yield for different energies.}
\begin{tabular}{|cccccc|cc|} \cline{1-8}
System&$E_{lab}$&$E^{*}$&$l_{cr}$&$l_{gr}$&NOC$^{a}$&\multicolumn{2}{|c|}{C yield (mb)} \\
&(MeV)&(MeV)&($\hbar$)&($\hbar$)&&expt&CASCADE$^{b}$  \\ \cline{1-8}
Ne + C&109.5&60&20&26&5.2&122.6$^{c}$&35.7 \\
Ne + C&145&73&24&31&4.5&151.9$\pm$26.8&62.2 \\
Ne + C&158&78&24&33&8.2&149.7$\pm$26.5&58.7 \\
Ne + C&170&82&24&34&11.0&131.8$\pm$23.8&56.9 \\
Ne + C&180&86&25&36&33.0&142.9$\pm$25.4&66.0 \\
Ne + C&200&94&25&38&73.2&178.3$\pm$30.7&51.4 \\ 
F + C&96&60&21&24&97.2&47.92$\pm$4.37 \cite{bhat96}&34.5 \\ \cline{1-8}
\end{tabular} 
\label{tbl1}
\end{table}
\noindent
$^{a}$ Values taken from Ref. \cite{beck94} \\
$^{b}$ Calculation done using $l_{cr}$ values given in Table~\ref{tbl1} \\
$^{c}$ Extracted from Fig.~\ref{nec7}

\end{document}